\documentclass{webofc}

\usepackage[varg]{txfonts}   
\usepackage{hyperref}
\usepackage{url}
\hypersetup{colorlinks=true,citecolor=blue,urlcolor=blue,linkcolor=blue}
\begin{document}
\title{Hydrodynamic effects on spin polarization along the beam direction in Au+Au and p+Pb collisions }
%
%

\author{\firstname{Cong} \lastname{Yi}\inst{1}\fnsep\thanks{{Speaker: congyi@mail.ustc.edu.cn}} \and
        \firstname{Xiang-Yu} \lastname{Wu}\inst{1,2}
        \and
        \firstname{Jie} \lastname{Zhu}\inst{1,3}
        \and
        \firstname{Shi} \lastname{Pu}\inst{4}
        \and
        \firstname{Guang-You} \lastname{Qin}\inst{1}
}

\institute{
Institute of Particle Physics and Key Laboratory of Quark and Lepton
Physics (MOE), Central China Normal University, Wuhan 430079, China  
\and Department of Physics, McGill University, Montreal, QC, Canada H3A
2T8
\and Fakult\"at f\"ur Physik, Universit\"at Bielefeld, D-33615 Bielefeld,
Germany
\and Department of Modern Physics, University of Science and Technology
of China, Hefei, Anhui 230026, China
          }

\abstract{
We investigate hydrodynamic effects on the spin polarization of $\Lambda$ hyperons in Au+Au collisions at $\sqrt{s_{NN}} = 200$ GeV and p+Pb collisions at $\sqrt{s_{NN}} = 8.16$ TeV using the CLVisc hydrodynamic framework. We present numerical results for the second Fourier sine coefficient of the longitudinal spin polarization, $\langle P_{z} \sin 2(\phi_{p} - \Psi_{2}) \rangle$, as a function of multiplicity (centrality) under three equilibrium scenarios: $\Lambda$ equilibrium, $s$-quark equilibrium, and isothermal equilibrium. We highlight the respective roles of thermal vorticity and the thermal-shear tensor in generating $\langle P_{z} \sin 2(\phi_{p} - \Psi_{2}) \rangle$ across collision systems and scenarios.}
\maketitle
\section{Introduction}
\label{intro}
The huge angular momentum and transverse anisotropic expansion generated
in peripheral heavy ion collisions can lead to global polarization \citep{Liang:2004ph,STAR:2017ckg}
and spin polarization along the beam direction \citep{STAR:2019erd,Becattini:2017gcx},
respectively. It has been found that in addition to thermal vorticity,
other hydrodynamic effects including the shear viscous tensor and
gradient of chemical potential can also contribute to local spin polarization
 \citep{Hidaka:2017auj,Becattini:2021suc,Liu:2021uhn}. These hydrodynamic
contributions to spin polarization have been extensively investigated
in nucleus-nucleus collisions \citep{Fu:2021pok,Becattini:2021iol,Yi:2021ryh,Alzhrani:2022dpi,Wu:2022mkr}.
Recently, the CMS collaboration has measured spin polarization along
the beam direction in p+Pb collisions at $\sqrt{s_{NN}}=8.16$ TeV  \citep{CMS:2025nqr}
. However, theoretical calculations of spin polarization for small
collision systems remain limited. In this work, we extend the study
of these hydrodynamic effects to small collision systems and compare
their roles across different collision systems.

\section{Theoretical Framework}
\label{sec-1}
Based on the theoretical approaches, such as the Kubo formula \cite{Fu:2021pok}, quantum kinetic theory \citep{Yi:2021ryh}
or quantum statistics model \citep{Becattini:2021iol}, the fermion
spin polarization vector $\mathcal{S}^{\mu}(\mathbf{p})$ at local
thermal equilibrium can be decomposed as $\mathcal{S}^{\mu}(\mathbf{p})=\mathcal{S}_{\text{thermal }}^{\mu}(\mathbf{p})+\mathcal{S}_{\text{th-shear  }}^{\mu}(\mathbf{p})$
in absence of electromagnetic field and baryon chemical potential.
Here the $\mathcal{S}_{\text{thermal }}^{\mu}$and $\mathcal{S}_{\text{th-shear}}^{\mu}$ are the spin polarization induced by the thermal vorticity $\varpi_{\alpha\beta} =\frac{1}{2}\left[\partial_{\alpha} \left(\frac{ u_{\beta}}{T} \right) - \partial_{\beta}\left(\frac{ u_{\alpha}}{T} \right)\right]$ and the thermal-shear tensor, $\xi_{\alpha\beta} = \left[\partial_{\alpha} \left(\frac{ u_{\beta}}{T} \right) + \partial_{\beta}\left(\frac{ u_{\alpha}}{T} \right)\right]$, respectively.

To better capture the bulk properties of different collision systems, we adopt AMPT and Trento-3D initial conditions for $\sqrt{s_{NN}}=200$ GeV Au+Au and $\sqrt{s_{NN}}=8.16$ TeV p+Pb collisions, respectively. We then employ the (3+1)-D hydrodynamic model CLVisc \citep{Pang:2012he,Wu:2021fjf} to simulate the evolution of the collision system and obtain the hydrodynamic gradient profiles on the freeze-out hypersurface. The subsequent hadronic scattering process is described by the SMASH model \citep{SMASH:2016zqf}.  The parameters used in our simulations can successfully reproduce the multiplicity and $v_2$ of charged particles. Assuming that the hadronic scattering has little effect on spin polarization, we use the modified Cooper-Frye formula to calculate the second Fourier coefficient of the spin polarization along the beam direction, $\langle P_{z}\sin(2\phi-2\Psi_{2})\rangle$, in the $\Lambda$'s rest frame.
Additionally, we consider three widely used scenarios: $\Lambda$ equilibrium, $s$-quark equilibrium, and iso-thermal equilibrium. In the numerical simulation, we use the particle masses $m_{\Lambda}=1.116~\mathrm{GeV}$ and $m_{s}=0.3~\mathrm{GeV}$ in the modified Cooper-Frye formula for the $\Lambda$ equilibrium and $s$-quark equilibrium scenarios, respectively. In the iso-thermal equilibrium scenario, all temperature gradients $\partial_{\mu}T$ are neglected.  More details can be found in our previous works \citep{Yi:2021ryh,Yi:2024kwu}.

\section{Numerical Results}
\label{sec-2}
\begin{figure}[t]
\centering
\sidecaption
\includegraphics[width=7.5cm,clip]{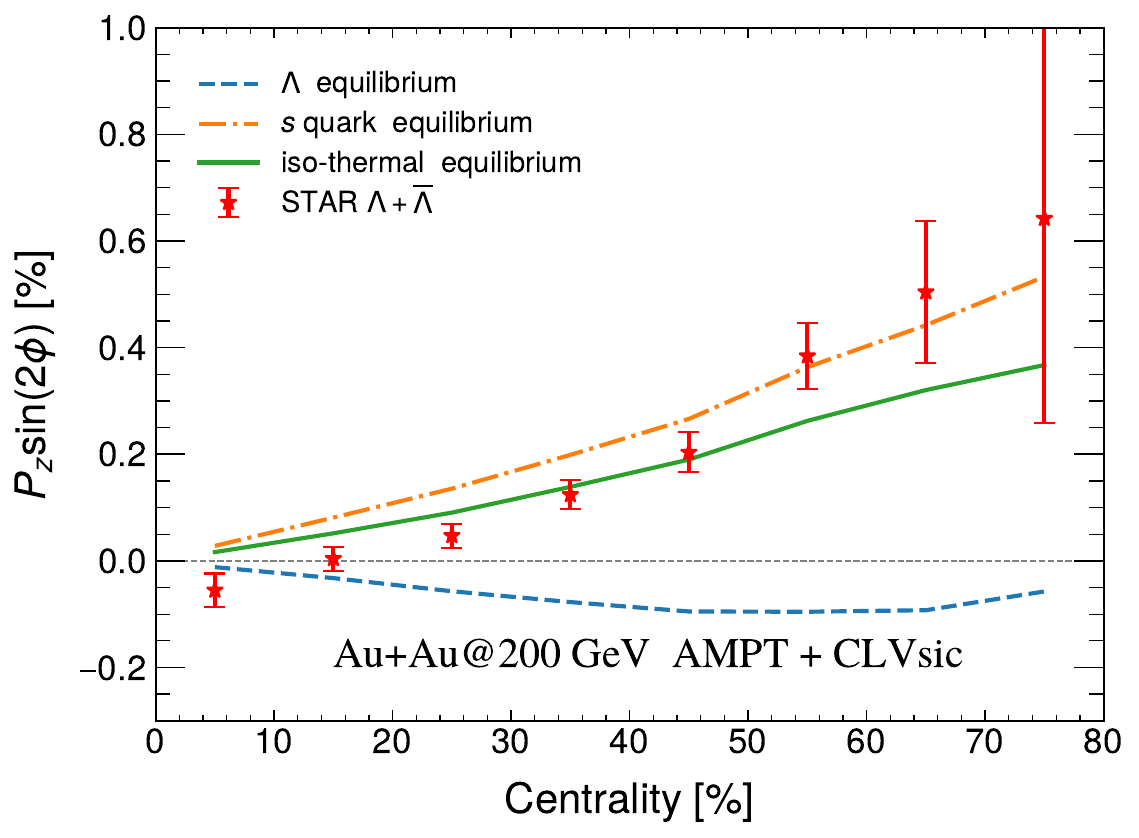}
\caption{The centrality dependence of $\langle P_{z}\sin(2\phi_{p}-2\Psi_{2})\rangle$ for $\Lambda$ hyperons in $\sqrt{s_{NN}} = 200$ GeV Au+Au collisions. 
Results are shown for three scenarios: $\Lambda$ equilibrium (blue dashed line), $s$ quark equilibrium (orange dash-dotted line), 
and iso-thermal equilibrium (green solid line). Red markers represent experimental data from Ref. \citep{STAR:2019erd}.}  
\label{fig:P2z-Au}       
\end{figure}

We present our numerical results for $\langle P_{z}\sin(2\phi_{p}-2\Psi_{2})\rangle$, under three equilibrium scenarios at $\sqrt{s_{NN}} = 200$ GeV Au+Au collisions in Fig.~\ref{fig:P2z-Au}. Our calculations indicate  that results for $s$-quark equilibrium (orange, dash-dot line) and iso-thermal equilibrium (green, solid line) agree with experimental data, while those for $\Lambda$-equilibrium (blue, dashed line) are inconsistent with measurements. These findings align with the results in Refs.~\citep{Fu:2021pok,Becattini:2021iol}. 

\begin{figure}[t]
\centering
\includegraphics[width=11cm,clip]{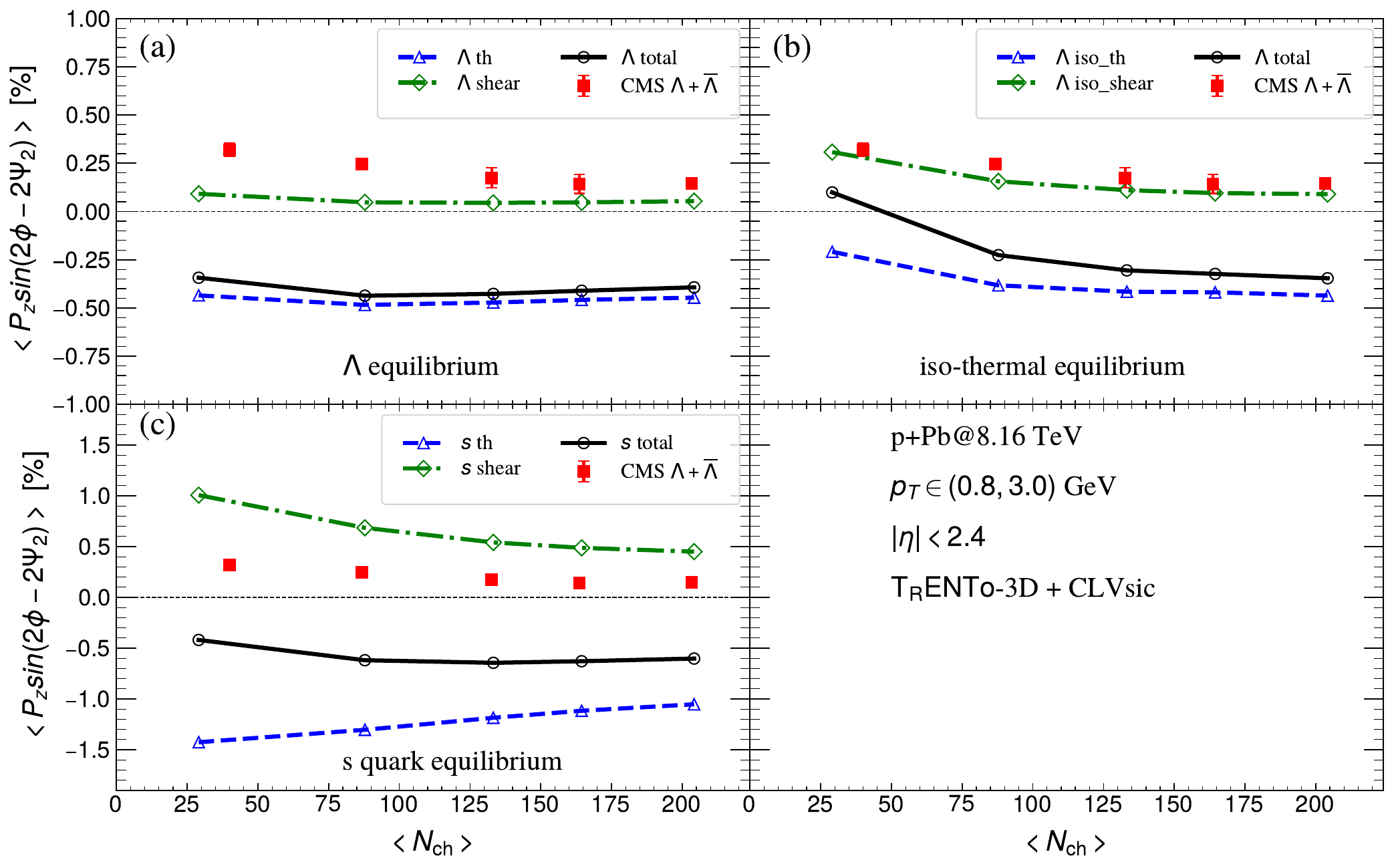}
\caption{The multiplicity $\langle N_{\rm ch}\rangle$ dependence of $\langle P_{z}\sin(2\phi_{p}-2\Psi_{2})\rangle$
for $\Lambda$ hyperons at $\sqrt{s_{NN}}$ = 8.16 TeV p+Pb collisions. Results are shown for three scenarios: $\Lambda$ equilibrium (a)
, $s$ quark equilibrium (b), and iso-thermal equilibrium (c). The contributions from thermal vorticity (blue triangles), thermal-shear tensor (green diamonds), and their combined effects (black circles) are presented. Red points correspond to CMS experimental data \citep{CMS:2025nqr}.}  
\label{fig:P2z_muliti}       
\end{figure}
In Fig. \ref{fig:P2z_muliti}, we show the charged-particle multiplicity
$\langle N_{\mathrm{ch}}\rangle$ dependence of $\langle P_{z}\sin(2\phi_{p}-2\Psi_{2})\rangle$
for $\Lambda$ hyperons in p+Pb collisions at $\sqrt{s_{NN}}=8.16$ TeV
under three scenarios. We observe that the shear-induced polarization
provides a positive contribution to the $\langle P_{z}\sin(2\phi_{p}-2\Psi_{2})\rangle$
, while thermal vorticity consistently yields a negative contribution
that opposes the shear term. These behaviors are consistent with those observed in Au+Au collisions. However, the competition between these two
mechanisms in p+Pb collisions produces an overall negative $\langle P_{z}\sin(2\phi_{p}-2\Psi_{2})\rangle$
across all scenarios. This net negative signal contradicts available
experimental data \citep{CMS:2025nqr}, revealing a significant discrepancy
between theoretical calculation and measurements in small collision
systems. These findings indicate that current hydrodynamic effects
alone cannot explain the spin polarization in p+Pb collisions, highlighting
the need for new polarization mechanisms in small systems for the future
research. 

It should be noted that our polarization results at lowest multiplicities are unreliable due to the limitations of hydrodynamic applicability. Nevertheless, our model can successfully describe the multiplicities and $v_{2}$ in the high-multiplicity regime. Therefore, we provide more detailed discussions on the beam-direction spin polarization as functions of azimuthal angle and rapidity for high-multiplicity p+Pb events in Ref.~\citep{Yi:2024kwu}, which are not presented here due to space constraints.

\section{Summary}
\label{sec-3}
In this work, we have employed the relativistic hydrodynamic model CLVisc to investigate the second Fourier coefficient of the beam-direction spin polarization of $\Lambda$ hyperons in both Au+Au and p+Pb collisions. We found that the shear-induced polarization consistently provides a positive contribution to $\langle P_{z}\sin(2\phi_{p}-2\Psi_{2})\rangle$, while the thermal vorticity effect always yields a negative contribution in both collision systems. The combined effects of hydrodynamic gradients can successfully describe the centrality dependence of the beam-direction spin polarization for $\Lambda$ hyperons in  $\sqrt{s_{NN}}=200$ GeV Au+Au collisions.  However, these hydrodynamic effects fail to account for the spin polarization data in p+Pb collisions. Our findings imply that the spin polarization in small collision systems cannot be explained solely by current hydrodynamic mechanisms and new physical mechanisms need to be explored in future studies.

\section*{Acknowledgements:} 
This work is supported in part by the National Key Research and Development
Program of China under Contract No. 2022YFA1605500, by the Chinese
Academy of Sciences (CAS) under Grants No. YSBR-088 and by National Natural Science Foundation of China (NSFC) under Grant Nos. 12075235,
12225503, 11890710, 11890711, 11935007, 12175122, 2021-867, 11221504,
11861131009 and 11890714.  X-Y.W was supported in part by the Natural Sciences and Engineering Research Council of Canada  (NSERC) [SAPIN-2020-00048 and SAPIN-2024-00026], and in part by US National Science Foundation (NSF) under grant number OAC-2004571. J.Z. was supported in part by China Scholarship Council (CSC) under Grant No. 202306770009.

\end{document}